\documentclass[conference,letterpaper]{IEEEtran}
\IEEEoverridecommandlockouts
\usepackage{cite}
\usepackage{amsmath,amssymb,amsfonts}
\usepackage{algorithmic}
\ifCLASSINFOpdf
\usepackage[pdftex]{graphicx}
\graphicspath{{./images/}}
\DeclareGraphicsExtensions{.pdf,.jpeg,.png}
\else
\usepackage[dvips]{graphicx}
\graphicspath{{./images/}}
\DeclareGraphicsExtensions{.eps}
\fi
\usepackage{textcomp}
\usepackage{xcolor}
\usepackage{array}
\def\BibTeX{{\rm B\kern-.05em{\sc i\kern-.025em b}\kern-.08em
    T\kern-.1667em\lower.7ex\hbox{E}\kern-.125emX}}

\DeclareMathOperator*{\argmax}{argmax}

\newcommand\rv[1]{{\color{black}#1}}
\newcommand\stecer[1]{{\color{black}#1}}

\begin{document}

\title{{Efficient Cooperative HARQ for Multi-Source Multi-Relay Wireless Networks}\\
}

\author{\IEEEauthorblockN{1\textsuperscript{st} Stefan~Cerovi{\'c}}
\IEEEauthorblockA{\textit{Orange Labs}\\
Ch{\^a}tillon, France \\
stefan.cerovic@orange.com}
\and
\IEEEauthorblockN{2\textsuperscript{nd} Rapha{\"e}l Visoz}
\IEEEauthorblockA{\textit{Orange Labs}\\
Ch{\^a}tillon, France \\
raphael.visoz@orange.com}
\and
\IEEEauthorblockN{3\textsuperscript{rd} Louis Madier}
\IEEEauthorblockA{\textit{Nokia}\\
Nozay, France \\
louis.madier@nokia.com}
}

\maketitle

\IEEEoverridecommandlockouts
\IEEEpubid{\makebox[\columnwidth]{978-1-5386-5541-2/18/\$31.00~\copyright2018 IEEE \hfill} \hspace{\columnsep}\makebox[\columnwidth]{ }}
\maketitle
\IEEEpubidadjcol

\begin{abstract}
In this paper, we compare the performance of three different cooperative Hybrid Automatic Repeat reQuest (HARQ) protocols for slow-fading half-duplex orthogonal multiple access multiple relay channel. Channel State Information (CSI) is available at the receiving side of each link only. Time Division Multiplexing is assumed, where each orthogonal transmission occurs during a time-slot. Sources transmit in turns in consecutive time slots during the first transmission phase. During the second phase, the destination schedules in each time-slot one node (source or relay) to transmit redundancies based on its correctly decoded source messages (its decoding set) with the goal to maximize the average spectral efficiency. Bidirectional limited control channels are available from sources and relays towards the destination to implement the necessary control signaling of the HARQ protocols. Among the three proposed HARQ, two follow the Incremental Redundancy (IR) approach. One consists in sending incremental redundancies on all the messages from the scheduled node decoding set (Multi-User encoding) while the other one helps a single source (Single User encoding) chosen randomly. The third one is of the Chase Combining (CC) type, where the selected node repeats the transmission (including modulation and coding scheme) of one source chosen randomly from its decoding set. Monte-Carlo simulations confirm that the IR-type of HARQ with Multi-User encoding offers the best performance, followed by IR-type of HARQ with Single User encoding and CC-type of HARQ. We conclude that IR-type of HARQ with Single User encoding offers the best trade-off between performance and complexity for a small number of sources in our setting.
\end{abstract}

\begin{IEEEkeywords}
chase combining, incremental redundancy, HARQ, multi-source multi-relay wireless network, spectral efficiency
\end{IEEEkeywords}

\section{Introduction}
Cooperative diversity by using relays in wireless networks allows increasing the total throughput of the network while (possibly) relying on single antenna nodes. Fundamental principles and the main idea of cooperative communications can be found in \cite{b1}, where a three-terminal Relay Channel 
is studied. If a limited feedback control channel is available from the destination to the relaying nodes, the throughput can be increased by using a cooperative version of Hybrid Automatic Repeat reQuest (HARQ) protocol \cite{b2}. We investigate the performance of different flavors of cooperative HARQ protocols for the Multiple Access Multiple Relay Channel (MAMRC), denoted by ($M$,$L$,$1$)-MAMRC where $M$ is the number of sources, $L$ the number of relays. User cooperation is included in our model which means that the number of relays that can help a given source is $L+M-1$. Each source listens to the other node transmissions with the final goal to maximize its number of correctly decoded source messages. The performance metric is the average spectral efficiency. Transmissions are orthogonal in time. During the first phase, each source transmits in turn its message in consecutive time slots. During the second phase (retransmission phase) the destination schedules a relay or a source to transmit for each time slot. All nodes are half-duplex, i.e., they can not transmit and receive at the same time. All the links are subject to slow-fading and Additive White Gaussian Noise (AWGN). For that reason, we use the outage information-theory tool to analyze the performance of the different protocols. Each node can cooperate with its successfully decoded source messages or decoding set. Indeed, contrary to the classical Decode and Forward (DF) approach where a relay need to wait until it decodes all the source messages correctly, here, a node can cooperate as soon as its decoding set is not empty. This relaying behavior is called Selective Decode-and-Forward (SDF) relaying function. By receiving a Channel Distribution Information (CDI) from all sources and relays (average SNR of all links), the destination can perform slow-link adaptation. It consists in allocating a rate among a discrete Modulation and Coding Scheme (MCS) family to each source in order to maximize the average spectral efficiency. For each possible $M$-tuple of source rates, the optimal slow link adaption algorithm exhaustively check which one achieves the best metric by performing Monte-Carlo simulations over a sufficient number of channel outcomes. The rate allocation is conveyed by a slow limited control channel from destination to sources prior to the HARQ protocol. Note that the optimal algorithm based on exhaustive search for finding the $M$-tuple of rates is quickly becoming intractable for either a large MCS family and/or an increasing number of sources. We designed a low complexity search algorithm for these cases whose performance gets very close to the optimal one. Its detailed presentation is out of the scope of this paper. In the following, we always assume that a slow link adaption rate allocation takes place before any source transmission. Since the CDI variations are much slower than the channel variations, the slow link adaptation keeps valid for many channel outcomes (actually our simulation are performed for a fixed CDI). There exist a limited feedback broadcast control channel from the destination towards sources and relays (e.g., to carry the scheduling decision of the destination) and multiple unicast forward coordination control channels from sources and relays towards the destination (to help the destination to take its scheduling decision). Particular care is paid to minimize the control overhead in this paper. Among the three proposed HARQ protocols, one consists in sending incremental redundancies on all the messages \rv{from the scheduled node decoding set} (Multi-User encoding) while the other one helps a single source (Single User encoding) chosen randomly. The latter is particularly attractive since its implementation can reuse state-of-art rate compatible punctured codes such as low density parity check codes or turbo codes. The third one is of the chase combining (CC) type, where the selected node repeats the transmission (including modulation and coding scheme) of one source chosen randomly \stecer{from} its decoding set. It allows Maximal Ratio Combining (MRC) at the destination of all the transmissions related to a given source. We can expect that such a protocol behaves poorly in general compared to the IR-type of HARQ. 

In \cite{b3} and \cite{b4}, the performance of different HARQ protocols is investigated for the single source, single relay and single destination case. Both CC and IR types of HARQ protocol were analyzed. In \cite{b4}, it is shown that IR-type of HARQ performs better than CC-type of HARQ in terms of system outage probability, average number of retransmissions and average transmission rates. The advantage of using relay selection (with limited feedback) over distributed space-time block coded transmissions in multiple relay networks is shown in \cite{b5}. User co-operation is included as in our paper. In \cite{b6}, Multi-User relay channel consisting of two sources, one relay and a destination is shown to take benefit from Multi-User encoding (network coding). In this work, a feedback channel is assumed to be available from both the relay and the destination towards the sources. For the multiple-source multiple-relay channel, a relay ordering algorithm based on finite field network coding has been proposed in \cite{b7}. An outage analysis has been done for that protocol, where Separate Network Channel Coding (SNCC) is used in combination with the DF relaying protocol. In \cite{b8}, the relay selection strategies that aim to maximize the long-term aggregate throughput are studied for slow-fading MAMRC, where SDF relaying protocol is applied under the JNCC/JNCD framework (Multi-User encoding at the relays with Multi-User iterative joint decoding at the destination). A proper comparison between the two IR-type of HARQ and the CC-type of HARQ has not been performed by the previously mentioned works. Our goal in this paper is to identify the most efficient cooperative HARQ protocol, i.e., the one that achieves the best complexity-performance tradeoff keeping in mind that Single User encoding and decoding is well mastered in terms of code construction and, clearly, less complex than Multi-User encoding and iterative joint decoding. On the other hand, the Chase Combining approach can be considered as having a similar complexity \stecer{to} Single User IR-HARQ. As a result, the HARQ protocol comparison comes down to a performance comparison where information theory outage analysis is particularly relevant.  

The remainder of the paper is organized as follows. The system model is detailed in section \ref{sec:sys_model}. In section \ref{sec:harq} the performance metric, the outage event definitions, as well as the three different HARQ protocols together with the proposed node selection strategy are described. Numerical results are presented in section \ref{sec:numerical_results}. Finally, we conclude the paper in section \ref{sec:conclusion}.

\section{System Model}
\label{sec:sys_model}
\begin{figure}[!t]
\centering
\includegraphics[scale=0.3]{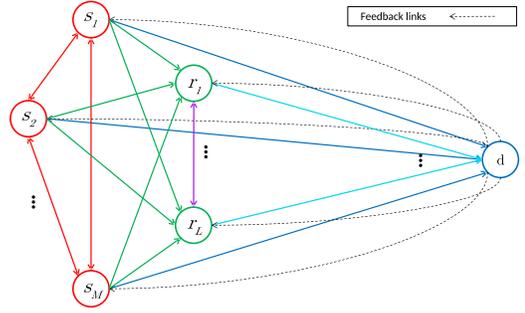}
\caption{Cooperative Orthogonal Multiple Access Multiple Relay Channel (OMAMRC) with feedback.}
\label{fig:fig_1}
\end{figure}

In this paper, we investigate OMAMRC under slow-fading assumption. $M$ sources, belonging to the set $\mathcal{S}=\{s_1,\dots,s_M\}$, transmit independent messages $\textbf{u}_s \in \mathbb{F}_2^{K_s}$ of $K_s$ information bits towards a common destination. The length of a source message depends on the selected MCS by the destination, where the decision about the selection is conveyed over the error-less limited feedback broadcast control channel. $L$ relays, that operate in half-duplex mode and that belong to the set $\mathcal{R}=\{r_1,\dots,r_L\}$, help the destination in decoding the sources' messages. They overhear the messages from sources due to the broadcast property of wireless medium, and apply SDF relaying protocol. Relays do not have their own messages to transmit. Additionally, user-cooperation is performed, i.e. when not transmitting, sources listen to other sources and relays transmissions and help the decoding at the destination by applying the SDF relaying protocol (see Fig. \ref{fig:fig_1}). Moreover, HARQ protocol is used, which is either of type Incremental Redundancy, or Chase Combining. In the case of IR-type of HARQ protocol, two types of encoding are considered: \stecer{S}ingle \stecer{U}ser encoding and \stecer{M}ulti-\stecer{U}ser encoding, depending on the number of sources that the node performing the relaying functions will help during its transmission. We define the set of all source and relay nodes as $\mathcal{N}=\mathcal{S} \cup \mathcal{R}$.

CSI is available only at the receiver side of each link and is assumed perfect. Hence, the destination only has the perfect knowledge of CSI of source-to-destination (S-D) links, $\textbf{h}_{\textrm{S,D}}=[ h_{s_1,d},\dots,h_{s_M,d} ]$, and of relay-to-destination (R-D) links $\textbf{h}_{\textrm{R,D}}=[ h_{r_1,d},\dots,h_{r_L,d} ] $. On the other hand, the CSI of source-to-source (S-S), source-to-relay (S-R) and relay-to-relay (R-R) links are unknown to it.

\begin{figure}[!t]
\centering
\includegraphics[scale=0.3]{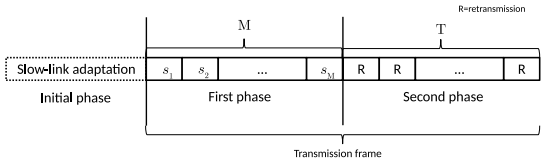}
\caption{Transmission of a frame: initial, first and second phase.}
\label{fig:fig_2}
\end{figure}

Transmission of source messages is split into frames, during which exactly one message from each source is sent, as well as the retransmissions related to those messages. Slow (block) fading is assumed, where within one frame the radio-links between the different nodes are considered to be fixed, while they change independently from frame to frame. Furthermore, we consider that during a certain number of frames $N_f>>1$, the probability distribution of the quality of each link remains constant. That means that the quality of the given link in the given frame represents one realization of the associated probability distribution. The choice of the MCS for each source by the destination takes place in the ``initial phase'' by applying the slow-link adaptation algorithm. That phase occurs before any transmission, and is repeated whenever the probability distributions of different channels change (see Fig. \ref{fig:fig_2}). CDI of each link in the network is needed as an input to the slow-link adaptation algorithm. For S-S, S-R and R-R links, sources and relays convey the information about CDI  to the destination over forward coordination control channels that are assumed to be errorless. The destination can track the CDI of S-D and R-D links by itself. The information about the selected MCSs is conveyed from the destination to all nodes over limited feedback  control channel. The source rates are kept fixed between two occurrences of the initial phase.

Transmission frame is split into two phases. The first phase consists of $M$ time-slots made of $N_1$ channel uses each, where each one of the $M$ sources transmits in turn. User co-operation being used, when one source transmits a message, both relays and non-transmitting sources listen and try to decode that message relying on a Cyclic Redundancy Check (CRC) code for error detection. The second phase consists of maximum of $T$ time-slots of duration of $N_2$ channel uses each, called also ``retransmission rounds'' in the following. $T$ is a system design parameter chosen by the destination, which depends on the latency requirements. In each retransmission round the destination selects one relay or a source to transmit, where a source can either retransmit its own message or act as a relay for other sources. In \cite{b8}, the scheduling strategy that consists in selecting the node whose link to the destination has the best quality among the nodes that can help the destination (their decoding sets contain at least one message that the destination has not been able to decode at the end of the previous round) is shown to achieve a performance close to the optimal (exhaustive) one. Taking into account the teaching of \cite{b8}, we propose a low overhead control signaling exchanges between the destination and the other nodes as follows:
\begin{itemize}
\item The destination broadcasts $M$ bits that indicate its decoding set $\mathcal{S}_{d,t-1}$ after round $t-1$ over the control channel.
\item If the decoding set of the destination consists of all source messages, a new frame begins and the sources transmit new messages while the relays and destination empty their memory buffers. Otherwise, each cooperating source and each relay which was able to decode at least one source message that is not included in the decoding set of the destination \rv{sends a signal on a dedicated unicast control channel. Each cooperating source or relay which did not decode any message needed by the destination, i.e., any message that is not included in the decoding set of the destination after round $t-1$, remains silent (ON-OFF modulation).}
\item Using the adopted node selection strategy, the destination can make the scheduling decision about the node to select for transmission. Its decision is broadcasted using a control channel.
\item Selected node transmits applying the appropriate type of HARQ protocol.
\end{itemize}

Note that the end of the first phase is considered as the end of the round zero. The non-selected nodes in a given retransmission round can benefit from the transmission of the scheduled node as well, and update their decoding sets accordingly. The number of retransmission rounds used in the second phase $T_{\mathit{used}} \in \{1,\dots,T\}$ depends on the success of the decoding process at the destination. Each node in the network is equipped with one antenna only and transmits with the same power. In the rest of the paper, the following notations are used:
\begin{itemize}
\item $x_{a,k}\in \mathbb{C}$ is the coded modulated symbol \rv{whose power is normalized to unity} for channel use $k$, sent from node $a\in \mathcal{S}\cup \mathcal{R}$.
\item $y_{a,b,k}$ is a received signal at node $b\in \mathcal{S}\cup \mathcal{R}\cup\{d\} \setminus \{a\}$, originating from node $a$.
\item $\gamma_{a,b}$ is the average signal-to-noise ratio (SNR) that captures both path-loss and shadowing effects.
\item $h_{a,b}$ are the channel fading gains, which are independent and follow a zero-mean circularly symmetric complex Gaussian distribution with variance $\gamma_{a,b}$.  
\item $n_{a,b,k}$ are independent and identically distributed AWGN samples, which follow a zero-mean circularly-symmetric complex Gaussian distribution with unit variance.
\end{itemize}

Using the previous notation, we can represent the received signal at node $b\in \mathcal{S}\cup \mathcal{R}\cup\{d\} \setminus \{a\}$ which originates from node $a\in \mathcal{S}\cup \mathcal{R}$ as:
\begin{equation}
y_{a,b,k}=h_{a,b}x_{a,k}+n_{a,b,k},
\end{equation}
where $k$ denotes a current channel use, taking a value $k \in \{1,\dots,N_1\}$ during the first phase, and $k \in \{1,\dots,N_2\}$ during the second phase.

\section{Cooperative HARQ protocols}
\label{sec:harq}
\subsection{Performance metric and outage events}
Let us denote with $R_s=K_s/N_1$ the initial transmission rate of a source $s$ in bit per complex dimension or bit per channel use [b.c.u]. We can define a long-term transmission rate $\bar{R}_s$ per source as the fraction of the number of transmitted information bits over the total number of channel uses spent, for a number of frames that tends to infinity:
\begin{equation}
\bar{R}_s=\frac{R_s}{M+\alpha\mathbb{E}(T_{\text{used}})},
\label{eq:first}
\end{equation}
where $\mathbb{E}(T_{\text{used}})=\sum_{t=1}^{T}t\textrm{Pr}\{T_{\text{used}}=t\}$ is the average number of retransmission rounds used in the second phase, and $\alpha=N_2 / N_1$.

A performance metric that we use throughout the paper is the average spectral efficiency, which can be defined as:
\begin{equation}
\eta=\sum_{i=1}^{M}\bar{R}_{s_i}(1-\textrm{Pr}\{\mathcal{O}_{s_i,T}\}),
\label{eq:third}
\end{equation}
where $\mathcal{O}_{s,T}$ is the ``individual outage event of source $s$ after round $T$'', which is the event that source $s$ is not decoded correctly at the destination after  round $T$. 

Before defining it analytically for different HARQ protocols in the following subsections, we should emphasize that the individual outage event of source $s$ after round $t$, $\mathcal{O}_{s,t}(a_t,\mathcal{S}_{a_t,t-1} |\textbf{h}_{\textrm{S,D}},\textbf{h}_{\textrm{R,D}},\mathcal{P}_{t-1})$, directly depends on the choice of a transmitting node $a_t\in \mathcal{N}$ in round $t$ and its associated decoding set $\mathcal{S}_{a_t,t-1}$. Furthermore, it is conditional on the knowledge of $\textbf{h}_{\textrm{S,D}}$, $\textbf{h}_{\textrm{R,D}}$ and $\mathcal{P}_{t-1}$, the last one denoting the set which collects the nodes $\hat{a}_k$ selected in rounds $k \in \{ 1,\dots,t-1 \}$ prior to round $t$ together with their associated decoding sets $\mathcal{S}_{\hat{a}_k,k-1}$, and the decoding set of the destination $\mathcal{S}_{d,t-1}$. The same holds for the ``common outage event after round $t$'' $\mathcal{E}_t(a_t,\mathcal{S}_{a_t,t-1} |\textbf{h}_{\textrm{S,D}},\textbf{h}_{\textrm{R,D}},\mathcal{P}_{t-1})$, which is the event that at least one source is not decoded correctly at the destination at the end of the round $t$. If $\mathbb{E}\{.\}$ is the expectation operator, and $\textbf{1}_{\{\mathcal{V}\}}$ is the function having a value $1$ if the event $\mathcal{V}$ is true, and $0$ otherwise, we can define the probability of the individual outage event of source $s$ after round $t$ for a candidate node $a_t$ as $\mathbb{E}\{\textbf{1}_{\{\mathcal{O}_{s,t}(a_t,\mathcal{S}_{a_t,t-1} |\textbf{h}_{\textrm{S,D}},\textbf{h}_{\textrm{R,D}},\mathcal{P}_{t-1})\} } \}$. We can define the probability of the common outage event in a similar way. In order to simplify the notation in the rest of the paper, we will omit the condition on $\textbf{h}_{\textrm{S,D}}$, $\textbf{h}_{\textrm{R,D}}$ and $\mathcal{P}_{t-1}$ when \stecer{recalling} the individual and common outage events.

\subsection{IR-type of HARQ protocol with \stecer{M}ulti-\stecer{U}ser \stecer{en}coding}
\label{MU}
In this part, we assume that in given round $t$, the selected node $a_t$ sends incremental redundancies on all the messages in its decoding set. If the decoding set at the destination after round $t-1$ is given by $\mathcal{S}_{d,t-1}$, we define the set of non-successfully decoded sources at the destination as $\bar{\mathcal{S}}_{d,t-1}=\mathcal{S} \setminus {\mathcal{S}}_{d,t-1}$. First, we want to analytically define the common outage event $\mathcal{E}_{t,\mathcal{B}}^{\text{IR,MU}}(a_t,\mathcal{S}_{a_t,t-1})$ after round $t$ for a candidate node $a_t$ of some subset $\mathcal{B}$ of the set of non-successfully decoded sources at the destination $\mathcal{B}\subseteq \bar{\mathcal{S}}_{d,t-1}$. Since in a given round the transmitted incremental redundancies potentially contain multiple source messages, the destination has no choice but to decode the source messages jointly, i.e., considering the received transmissions as part of a joint codeword on all the source messages. As a result, we resort to Multiple Access Channel (MAC) framework, where the event $\mathcal{E}_{t,\mathcal{B}}^{\text{IR,MU}}(a_t,\mathcal{S}_{a_t,t-1})$ is true if the vector of rates of sources contained in $\mathcal{B}$ lies outside of the corresponding MAC capacity region.

We can express this event as:
\begin{equation}
\begin{split}
&\mathcal{E}_{t,\mathcal{B}}^{\text{IR,MU}}(a_t,\mathcal{S}_{a_t,t-1})=\bigcup_{\mathcal{U}\subseteq \mathcal{B}} \Big\{ \sum_{s \in \mathcal{U}}R_s > \sum_{s \in \mathcal{U}} I_{s,d}\\
&+ \sum_{l=1}^{t-1} \alpha I_{\hat{a}_l,d} \textbf{1}_{\{\mathcal{C}_{\hat{a}_l}^{\text{IR,MU}} \}} + \alpha I_{a_t,d} \textbf{1}_{\{ \mathcal{C}_{a_t}^{\text{IR,MU}} \}} \Big\},
\end{split}
\label{eq:reduced_mac1}
\end{equation}
where $I_{a,b}$ denotes the mutual information between the nodes $a$ and $b$, the sources contained in the set $\mathcal{I}=\bar{\mathcal{S}}_{d,t-1}\setminus \mathcal{B}$ are considered as interference and $\mathcal{C}_{\hat{a}_l}^{\text{IR,MU}}$ and $\mathcal{C}_{a_t}^{\text{IR,MU}}$ have the following definitions:
\begin{equation}
\label{eq:IR_MU_C_com}
\begin{split}
&\mathcal{C}_{\hat{a}_l}^{\text{IR,MU}}=\Big\{ \{\mathcal{S}_{\hat{a}_l,l-1}\cap \mathcal{U} \neq \emptyset\}\wedge \{\mathcal{S}_{\hat{a}_l,l-1}\cap \mathcal{I}=\emptyset\} \Big\},\\
&\mathcal{C}_{a_t}^{\text{IR,MU}}=\Big\{ \{\mathcal{S}_{a_t,t-1}\cap \mathcal{U} \neq \emptyset\}\wedge \{\mathcal{S}_{a_t,t-1}\cap \mathcal{I}=\emptyset\} \Big\},
\end{split}
\end{equation}
with $\wedge$ standing for the logical and. Since IR-type of HARQ is used, we basically compare the sum-rate of sources contained in each subset $\mathcal{U}\subseteq \mathcal{B}$ with the accumulated mutual information at the destination that originates from: (1) the transmissions during the first phase; (2) the transmissions of previously activated nodes in rounds $1, \dots, t-1$; and (3) the transmission of the candidate node $a_t$. Node $\hat{a}_k$ for $k=\{1,\dots,T\}$ is involved in the calculation only if it was able to successfully decode at least one source from the set $\mathcal{U}$ while its decoding set does not contain any interference, i.e., source message that \stecer{is} outside $\mathcal{B}$. Multiplication by $\alpha$ serves as a normalization before adding two mutual information originating from two different phases, where the transmission uses $N_1$ and $N_2$ time slot channel uses in the first and second phase, respectively. If one or more MAC inequalities associated to the sum-rate of sources in different sets $\mathcal{U}$ is not respected, the common outage event of the set $\mathcal{B}$ is proclaimed.

By similar reasoning, the individual outage event of the source $s$ after round $t$ can be defined as:
\begin{equation}
\begin{split}
&\mathcal{O}_{s,t}^{\text{IR,MU}}(a_t,\mathcal{S}_{a_t,t-1})=\bigcap_{\mathcal{I}\subset \bar{\mathcal{S}}_{d,t-1}} \bigcup_{\mathcal{U}\subseteq \bar{\mathcal{I}}:s\in \mathcal{U}}\Big\{ \sum_{s \in \mathcal{U}}R_s > \sum_{s \in \mathcal{U}} I_{s,d}\\
 &+ \sum_{l=1}^{t-1} \alpha I_{\hat{a}_l,d} \textbf{1}_{\{\mathcal{C}_{\hat{a}_l,s}\}} + \alpha I_{a_t,d} \textbf{1}_{\{\mathcal{C}_{a_t,s}\}} \Big\},
\label{eq:Ost_JNCC}
\end{split}
\end{equation}
where $\bar{\mathcal{I}}=\bar{\mathcal{S}}_{d,t-1}\setminus \mathcal{I}$, and $\mathcal{C}_{\hat{a}_l,s}^{\text{IR,MU}}$ and $\mathcal{C}_{a_t,s}^{\text{IR,MU}}$ have the following definitions:
\begin{equation}
\label{eq:IR_MU_C_ind}
\begin{split}
&\mathcal{C}_{\hat{a}_l,s}^{\text{IR,MU}}=\Big\{ \{s\in \mathcal{S}_{\hat{a}_l,l-1}\cap \mathcal{U}\}\wedge \{\mathcal{S}_{\hat{a}_l,l-1}\cap \mathcal{I}=\emptyset\} \Big\},\\
&\mathcal{C}_{a_t,s}^{\text{IR,MU}}=\Big\{ \{s\in \mathcal{S}_{a_t,t-1}\cap \mathcal{U}\}\wedge \{\mathcal{S}_{a_t,t-1}\cap \mathcal{I}=\emptyset\} \Big\},
\end{split}
\end{equation}

\subsection{IR-type of HARQ protocol with \stecer{S}ingle \stecer{U}ser encoding}
\label{SU}
As stated in the introduction, Single User encoding is particularly attractive since its implementation can reuse state-of-art rate compatible punctured codes such as low density parity check codes or turbo codes. Here, a selected node in retransmission round $t$ of the second phase cooperates with a single source from its decoding set, i.e., it transmits incremental redundancies for a single source. The choice of the source that the selected node will help is random, but among all sources which the destination has not successfully decoded up until that round. That information is available to each node in the network due to the control information exchange mechanism described in section \ref{sec:sys_model}. 

Let us denote with $s_{\hat{a}_k}$ a randomly chosen source by the node $\hat{a}_k$ in round $k \in \{1, \dots, T\}$ from its decoding set under the previously described condition. In this case, since the selected nodes during the second phase do not apply \stecer{M}ulti-\stecer{U}ser encoding anymore and since the transmission is orthogonal in time, there is no need to use the MAC framework. The individual outage event of the source $s$ after round $t$ for the selected node $a_t$ which cooperates with the source $s_{a_t}$ can be simply defined as:
\begin{equation}
\begin{split}
&\mathcal{O}_{s,t}^{\text{IR,SU}}(a_t,s_{a_t})=\Big\{ R_s >  I_{s,d} + \sum_{l=1}^{t-1} \alpha I_{\hat{a}_l,d} \textbf{1}_{\{s=s_{\hat{a}_l}\}}\\
&+ \alpha I_{a_t,d} \textbf{1}_{\{s=s_{a_t}\}} \Big\},
\label{eq:Ost_JNCC_SU}
\end{split}
\end{equation}
To find the common outage event of sources contained in the set $\mathcal{B}\subseteq \bar{\mathcal{S}}_{d,t-1}$ after round $t$, for the selected node $a_t$ which cooperates with the source $s_{a_t}$, we simply check if the individual outage event of any source $s$ contained in $\mathcal{B}$ is true:
\begin{equation} 
\mathcal{E}_{t,\mathcal{B}}^{\text{IR,SU}}(a_t,s_{a_t})=\bigcup_{s \in \mathcal{B}}  \mathcal{O}_{s,t}^{\text{IR,SU}}(a_t,s_{a_t}).
\label{eq:reduced_mac2}
\end{equation}

\subsection{CC-type of HARQ protocol}
\label{CC}
In this type of protocol, the selected node $a_t$ in round $t$ in the second phase apply the exact same MCS as source $s$ whose message is randomly selected from the decoding set of the destination $\bar{\mathcal{S}}_{d,t-1}$. It implies the constraint that $N_1=N_2$ or $\alpha=1$. At the destination, Maximal Ratio Combining (MRC) (at symbol or coded bit level) is used after each round in order to decode the message of a given source. \rv{By doing so, we obtain the highest achievable SNR, denoted $\gamma_{\text{MRC}}$, for a given source at the destination which is equal to the summation of individual SNRs from the previous rounds. This kind of protocol offers less complexity in decoding then the protocol based on \stecer{M}ulti-\stecer{U}ser encoding. The individual outage event of the source $s$ after round $t$ for the selected node $a_t$ which cooperates with the source $s_{a_t}$ is defined in this case as:
\begin{equation}
\mathcal{O}_{s,t}^{\text{CC}}(a_t,s_{a_t})=\Big\{ R_s >  I(\gamma_{\text{MRC}}(a_t,s_{a_t}))\Big\}
\end{equation}
where
\begin{equation}
\begin{split}
&\gamma_{\text{MRC}}(a_t,s_{a_t}) = |h_{s,d}|^2 + \sum_{l=1}^{t-1}|h_{\hat{a}_l,d}|^2 \textbf{1}_{\{s=s_{\hat{a}_l}\}} \\
&+ |h_{a_t,d}|^2 \textbf{1}_{\{s=s_{a_t}\}} 
\end{split}
\label{eq:Ost_JNCC_CC}
\end{equation}
}

The common outage event of sources contained in the set $\mathcal{B}\subseteq \bar{\mathcal{S}}_{d,t-1}$ after round $t$, for the selected node $a_t$ which cooperates with the source $s_{a_t}$ is, just as in the previous case, defined as:
\begin{equation} 
\mathcal{E}_{t,\mathcal{B}}^{\text{CC}}(a_t,s_{a_t})=\bigcup_{s \in \mathcal{B}}  \mathcal{O}_{s,t}^{\text{CC}}(a_t,s_{a_t}).
\label{eq:reduced_mac3}
\end{equation}

\subsection{Node selection strategy used during the second phase}
\label{node_selection}

In \cite{b8}, it is shown that the node selection strategy which offers the best trade-off between the performance and computational complexity is the one where in a given round of the second phase the node with the highest mutual information between itself and the destination is selected among all nodes that were able to decode at least one source from the set of non-successfully decoded sources at the destination after round $t-1$:
\begin{equation}
\hat{a}_t=\argmax_{a_t\in\mathcal{S}\cup\mathcal{R}}\{ I_{a_t,d} \textbf{1}_{\{ \bar{\mathcal{S}}_{d,t-1}\cap \mathcal{S}_{a_t,t-1} \neq \emptyset \}}\}.
\label{eq:strategy1}
\end{equation}
Namely, it is demonstrated by performing Monte-Carlo simulations that such a strategy performs close to the upper-bound given by the strategy based on the exhaustive search for the best activation sequence, which requires the knowledge of the CSI of each link in the network and is much more complex. By observing the expressions for the individual outage probability, it is clear that such a node selection strategy can be equally applied to the IR-type of HARQ with the \stecer{S}ingle \stecer{U}ser encoding and the CC-type of HARQ.

\section{Numerical Results}
\label{sec:numerical_results}
In this Section, we want to evaluate the performance of the three types of HARQ protocols described in Sections \ref{MU}, \ref{SU}, \ref{CC} in terms of the average spectral efficiency by performing Monte-Carlo simulations. The node selection strategy described in subsection \ref{node_selection} is assumed to be used in the second phase. Also, we assume the presence of \stecer{the optimal} slow-link adaptation algorithm \stecer{conditional on the chosen node selection strategy}. A discrete MCS family whose rates belong to $\{$0.5$,$1$,$1.5$,$2$,$2.5$,$3$,$3.5$\}$ [b.c.u] is used for the initial rates. Independent Gaussian distributed channel inputs are assumed (with zero mean and unit variance) where $I_{a,b}=\log_2(1+|h_{a,b}|^2)$. There are some other formulas that could be also used for the calculation of $I_{a,b}$which take into account, for example, discrete entries, finite length of the codewords, non-outage achieving \stecer{Multi-User encoding/iterative joint decoding} architectures etc. Although the calculation would be different for each type of HARQ protocol, the basic concept of \stecer{the} work would stay the same.

In the first part of the simulations, we consider ($3$,$3$,$1$)-OMAMRC with $\alpha^{\text{IR}}=0.5$ and $T^{\text{IR}}=4$ for IR-types of HARQ protocol, and $\alpha^{\text{CC}}=1$ and $T^{\text{CC}}=2$ for CC-type of HARQ protocol. The asymmetric link configuration is assumed, where the average SNR of each link is in the range $\{-15\textrm{dB}, \dots, 20\textrm{dB}\}$, where the source $s_1$ is set on purpose to be in the best propagation condition, while the source $s_3$ is in the worst one. Concretely, the network is configured as follows: (1) the average SNR of the links between source $s_1$ and each relay, as well as the link between source $s_1$ and the destination, is set to $\gamma$; 
(2) the average SNR of the links between source $s_2$ and each relay, as well as the link between source $s_2$ and the destination, is set to $\gamma-4\textrm{dB}$; 
(3) the average SNR of the links between source $s_3$ and each relay, as well as the link between source $s_3$ and the destination, is set to $\gamma-7\textrm{dB}$; 
(4) the average SNR of the links between all relays, as well as the links between each relay and the destination is set to $\gamma$; 
(5) the average SNR of the links between all sources are set according to the Tab. \ref{tab:tab1}. 

\begin{figure}[!t]
\centering
\includegraphics[scale=0.37]{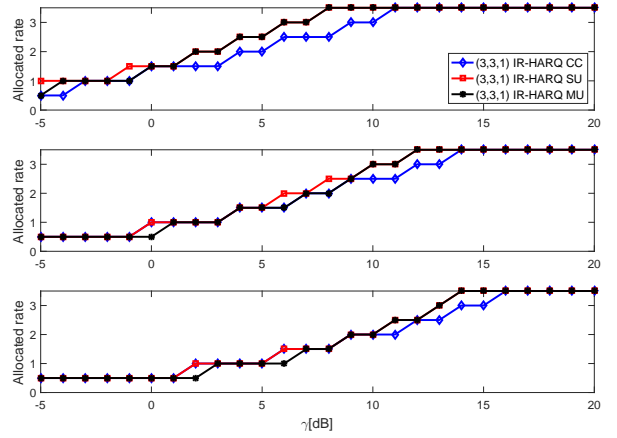}
\caption{Allocated rates to sources for different HARQ protocols for asymmetric link configuration in ($3$,$3$,$1$)-OMAMRC.}
\label{fig:figs1}
\end{figure}

\begin{figure}[!t]
\centering
\includegraphics[scale=0.37]{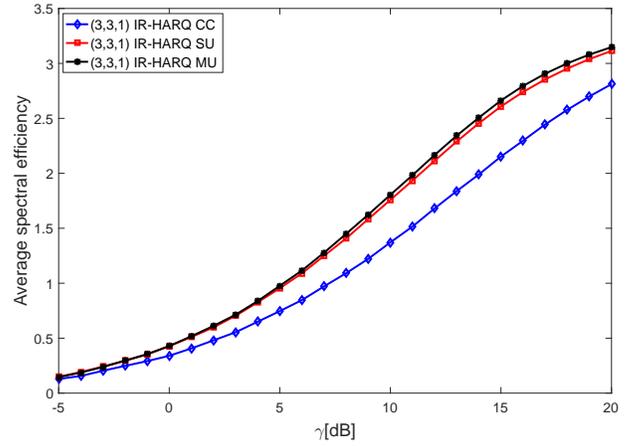}
\caption{Average spectral efficiency obtained by using different HARQ protocols for asymmetric link configuration in ($3$,$3$,$1$)-OMAMRC.}
\label{fig:figs2}
\end{figure}

\begin{table}[!t]
\caption{Average SNR of the links between all sources}
\label{tab:tab1}
\centering
\begin{tabular}{!{\vrule width 0.8pt} c !{\vrule width 0.8pt} c | c | c !{\vrule width 0.8pt}}
\noalign{\hrule height 0.8pt}
$\gamma_{x,y}[\textrm{dB}]$  & $s_1$ & $s_2$ & $s_3$\\
\noalign{\hrule height 0.8pt}
$s_1$ & N.A. & $\gamma-1\textrm{dB}$ & $\gamma-2\textrm{dB}$\\
\hline
$s_2$ & $\gamma-1\textrm{dB}$ & N.A. & $\gamma-5\textrm{dB}$\\
\hline
$s_3$ & $\gamma-2\textrm{dB}$ & $\gamma-5\textrm{dB}$ & N.A.\\
\noalign{\hrule height 0.8pt}
\end{tabular}
\end{table}
As a result, the initial rates associated to all sources are asymmetric. They are shown on Fig. \ref{fig:figs1} as a function of $\gamma$, which is the average SNR \stecer{of the link} between source $s_1$ and the destination. On that figure, IR-type of HARQ protocol with \stecer{M}ulti-\stecer{U}ser \stecer{en}coding is labeled as ``IR-HARQ MU'', IR-type of HARQ protocol with \stecer{S}ingle \stecer{U}ser \stecer{en}coding as ``IR-HARQ SU'' while CC-type of HARQ protocol is labeled as ``CC-HARQ''. Fig. \ref{fig:figs2} shows the average spectral efficiency of the network as a function of $\gamma$. We observe that the IR-type of HARQ protocol with \stecer{M}ulti-\stecer{U}ser \stecer{en}coding provides the highest average spectral efficiency. This result was expected since the selected nodes in the second phase may help the decoding of  multiple sources at the same time. IR-type of HARQ protocol with \stecer{S}ingle \stecer{U}ser \stecer{en}coding performance is not far behind, providing slightly lower average spectral efficiency. It can be explained by the fact that ``only'' three sources are present in the network, so there is often a case where the selected node in the second phase cooperates with exactly one source\stecer{, even if \stecer{M}ulti-\stecer{U}ser \stecer{en}coding is employed}. Naturally, CC-type of HARQ has a \stecer{noticeably worse} performance compared with two IR based protocols.  

\begin{figure}[!t]
\centering
\includegraphics[scale=0.37]{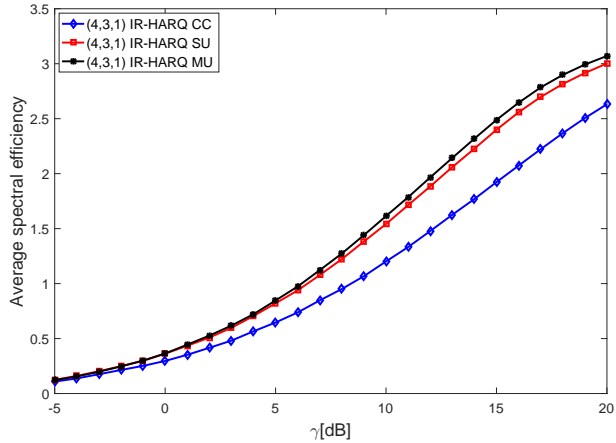}
\caption{Average spectral efficiency obtained by using different HARQ protocols for asymmetric link configuration in ($4$,$3$,$1$)-OMAMRC.}
\label{fig:figs3}
\end{figure}

\begin{figure}[!t]
\centering
\includegraphics[scale=0.37]{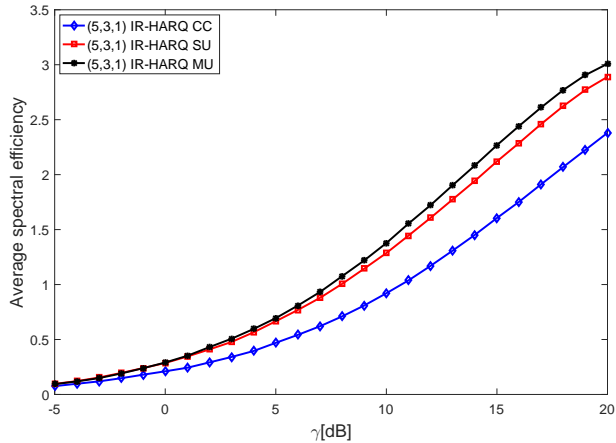}
\caption{Average spectral efficiency obtained by using different HARQ protocols for asymmetric link configuration in ($5$,$3$,$1$)-OMAMRC.}
\label{fig:figs4}
\end{figure}

Fig. \ref{fig:figs3} and Fig. \ref{fig:figs4} show the same comparison but for ($4$,$3$,$1$)-OMAMRC and ($5$,$3$,$1$)-OMAMRC, respectively. The average SNR of the links between source $s_4$ and each relay, as well as the link between source $s_4$ and the destination, is set to $\gamma-9\textrm{dB}$, while the average SNR of the links between source $s_5$ and each relay, as well as the link between source $s_5$ and the destination, is set to $\gamma-10\textrm{dB}$. The average SNR of the link between sources $s_4$ and $s_5$ is set to $\gamma-9.5\textrm{dB}$, while the same parameter for the links between sources $s_4$ and $s_5$ and all other sources is set to $\gamma$ reduced by a value from the set $[0\textrm{dB}, \dots, 9\textrm{dB}]$, following the similar logic as in the case of ($3$,$3$,$1$)-OMAMRC. We observe that the performance ordering of the different protocols remains the same. But, as the number of sources in the network grows,  we notice that \stecer{for IR-type of HARQ }the difference in performance between the \stecer{M}ulti-\stecer{U}ser and \stecer{S}ingle \stecer{U}ser \stecer{encoding} slowly grows. Indeed, a scheduled node has all the more chances to have more than one source in its decoding set as the number of sources increases.

As a general conclusion, we can argument that for the OMAMRC with relatively small number of sources the IR-type of HARQ with \stecer{S}ingle \stecer{U}ser \stecer{en}coding offers the best compromise between performance and complexity.

It is also interesting to observe that the average spectral efficiency decreases for all three types of the HARQ protocol when the number of sources increases. There are two reasons from our understandings. The first one is that by adding more sources that are progressively in worse conditions than the previous ones, the probability that the added source will be successfully decoded decreases. The other reason is that the number of retransmission rounds in the second phase is fixed to $T^{\text{IR}}=4$ and $T^{\text{CC}}=2$, so by adding more sources, even if they are all in the same conditions in average, \stecer{it may happens that }there are not enough available retransmissions for helping them all \stecer{efficiently}.

\begin{figure}[!t]
\centering
\includegraphics[scale=0.37]{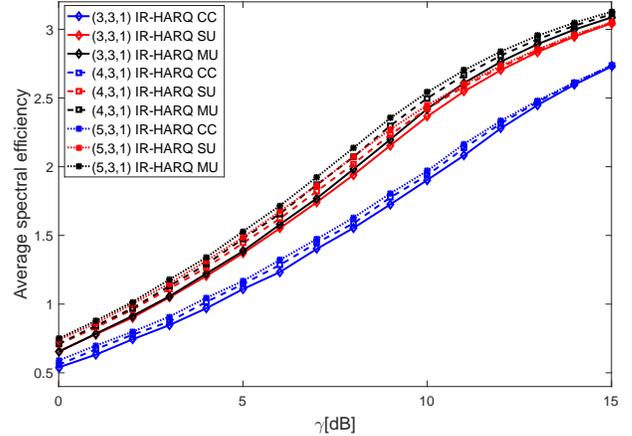}
\caption{Average spectral efficiency obtained by using different HARQ protocols for symmetric link configuration and different OMAMRC.}
\label{fig:figs5}
\end{figure}

For that reason, in the last part of simulations, we consider the symmetric link configuration where the average SNR of each link is equal to $\gamma$, and where the number of possible retransmission rounds in the second phase varies with the number of sources. Namely, we try to keep a constant ratio between the number of time-slots in the first phase and the number of possible retransmissions in the second phase. Let $M_1=3$ be the number of sources in ($3$,$3$,$1$)-OMAMRC, with $T_1^{\text{IR}}=4$ the number of retransmissions in the second phase for IR-type of HARQ, and with $T_1^{\text{CC}}=2$ the same number, but for CC-type of HARQ. In ($4$,$3$,$1$)-OMAMRC, $M_2=4$, $T_2^{\text{IR}}=\lceil \frac{T_1}{M_1}M_2 \rceil=6$, and $T_2^{\text{CC}}=\frac{T_2^{\text{IR}}}{2}=3$. In the case of ($5$,$3$,$1$)-OMAMRC, by similar reasoning and forcing the $T_3^{\text{CC}}$ to be the round number, we choose $T_3^{\text{IR}}=8$ and $T_3^{\text{CC}}=4$. Fig. \ref{fig:figs5} shows the comparison of the average spectral efficiency for all $M_1$, $M_2$ and $M_3$ where we see that in this case the more sources there are in the network, the higher the average spectral efficiency is. For the clarity of the figure only the range $\gamma \in \{0\textrm{dB}, \dots, 15\textrm{dB}\}$ is shown.

\section{Conclusion}
\label{sec:conclusion}
In this paper, we investigate the performance of three different cooperative HARQ protocols for slow-fading MAMRC. Among the three proposed HARQ protocols, two follow the \stecer{I}ncremental \stecer{R}edundancy (IR) approach. One consists in sending incremental redundancies \rv{on all the source messages decoded correctly by the scheduled node} (Multi-User encoding) while the other one helps a single source (Single User encoding) chosen randomly. The third one is of the \stecer{C}hase \stecer{C}ombining (CC) type, where the selected node repeats the transmission (including modulation and coding scheme) of one source chosen randomly in its correctly decoded source message set (its decoding set). It allows Maximal Ratio Combining (MRC) at the destination of all the transmissions related to a given source. Single User encoding and decoding is well mastered in terms of code construction (state of the art rate compatible punctured codes) and, clearly, less complex than Multi-User encoding and iterative joint decoding. On the other hand, the Chase Combining approach can be considered as having a similar complexity \stecer{to} Single User IR-HARQ. To identify the most efficient cooperative HARQ protocol, i.e., the one that achieves the best complexity-performance tradeoff, we resort to \rv{information theory outage based average spectral efficiency performance comparison}. We conclude that IR-type of HARQ with \stecer{S}ingle \stecer{U}ser encoding offers the best trade-off between performance and complexity for a small number of sources in our setting.

\vspace{12pt}


\begin{thebibliography}{00}
\bibitem{b1} E. C. Van Der Meulen, ``Three-terminal communication channels,'' \textit{Adv. Appl. Probab.}, vol. 3, no. 1, p. 120-154, 1971.
\bibitem{b2} C. Lott, O. Milenkovic and E. Soljanin, ``Hybrid ARQ: theory, state of the art and future directions,'' \textit{Proc. IEEE Inf. Theory Workshop Inf. Theory Wireless Netw.}, Solstrand, Norway, Jul. 2007.
\bibitem{b3} B. Makki, T. Eriksson and T. Svensson, ``On the performance of the relay–ARQ networks,'' \textit{IEEE Trans. Veh. Technol.}, vol. 65, no. 4, pp. 2078-2096, Apr. 2016.
\bibitem{b4} A. Chelli and M. S. Alouini, ``On the performance of hybrid-ARQ with incremental redundancy and with code combining over relay channels,'' \textit{IEEE Trans. Wireless Commun.}, vol. 12, no. 8, pp. 3860-3871, Aug. 2013.
\bibitem{b5} E. Beres and R. Adve, ``On selection cooperation in distributed networks,'' \textit{Proc. 2006 40th Annu. Conf. Inf. Sciences and Systems}, Princeton, NJ, USA, Mar. 2006.
\bibitem{b6} Y. Sun, Y. Li, and X. Wang, ``Cooperative hybrid-arq protocol with network coding,'' in  \textit{Proc. IEEE ChinaCOM 2009}, Xian, China, Aug. 2009.
\bibitem{b7} Y. Cheng and L. Yang, ``Joint relay ordering and linear finite field network coding for multiple-source multiple-relay wireless sensor networks,'' \textit{Int. J. Distrib. Sens. N.}, vol. 9, no. 10, pp. 1–12, 2013.
\bibitem{b8} S. Cerovic, R. Visoz, L. Madier, and A. O. Berthet, ``Centralized scheduling strategies for cooperative harq retransmissions in multi-source multi-relay wireless networks,'' \textit{Proc. IEEE ICC'18}, Kansas City, MO, USA, May 2018.
\end{thebibliography}
\end{document}